\documentclass[aip,apl,reprint,superscriptaddress,showpacs,10pt]{revtex4-1}

\usepackage{graphicx}
\usepackage{amsmath}
\usepackage{amstext}
\usepackage{amssymb}
\usepackage{color}
\begin{document}
\allowdisplaybreaks
% ================================================================================
% 
\title{Localizing energy in granular materials.}

% ================================================================================
%
\author{Michelle A. Przedborski}
\email{mp06lj@brocku.ca}
\author{Thad A. Harroun}
\email{thad.harroun@brocku.ca}
\affiliation{Department of Physics, Brock University, St. Catharines, Ontario, Canada L2S 3A1 }
\author{Surajit Sen}
\email{sen@buffalo.edu}
\affiliation{Department of Physics, State University of New York, Buffalo, New York 14260-1500, USA}

% ================================================================================
\date{\today}

% ================================================================================
%
%
\begin{abstract}
A device for absorbing and storing short duration impulses in an initially uncompressed one-dimensional granular chain  is presented. Simply stated, short regions of sufficiently soft grains are embedded in a hard granular chain. These grains exhibit long-lived standing waves of predictable frequencies regardless of the timing of the arrival of solitary waves from the larger matrix. We explore the origins, symmetry, and energy content of the soft region and its intrinsic modes.
\end{abstract}

% ================================================================================
%\keywords{}

% ================================================================================
\maketitle

% ================================================================================

Methods for trapping energy in granular materials has been a topic of interest to the scientific community in recent years. Controlling energy localization processes in such materials affords the opportunity to develop new energy-harvesting or filtering devices. One direction of this research is into the formation of intrinsic localized modes (or discrete breathers)~\cite{Campbell2004,Flach2008}, and there have been several experiments and simulated demonstrations of these excitations in one-dimensional granular chains~\cite{Job2009, Theocharis2009, Theocharis2010}. Other work has focused on 
energy trapping in granular containers~\cite{Hong2005,Vergara2006} and composite granular systems~\cite{Daraio2006energy,Breindel2011}.

Recently, we showed that softer grains in a monoatomic Hertzian chain results in slowing solitary wave (SW) propagation, and that soft boundaries causes a significant delay in the reflection of SWs due to the introduction of an inertial mismatch at the boundaries of the system~\cite{Przedborski2015}. This leads to the straightforward idea of localizing both kinetic and potential energy on a defect of soft and small grains, in a manner distinct from previous studies.

\begin{figure}[htp]
\centering
%\framebox(150,125){TEXT}
\includegraphics[width=0.45\textwidth]{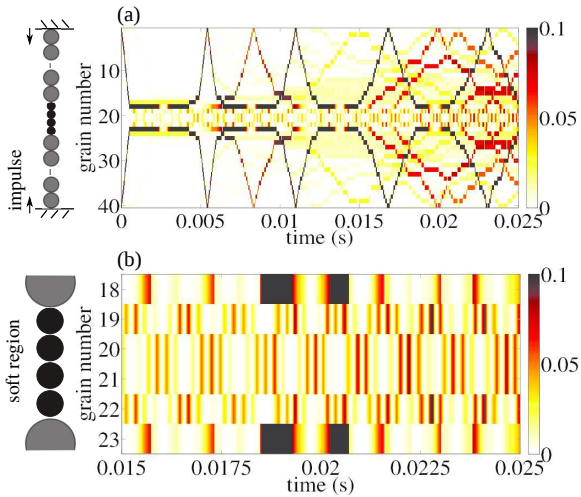}
\caption{Kinetic energy density plots for the $N=40$ granular chain with four central soft grains.  (a) shows how the kinetic energy is maintained in the central region long after the initial perturbation to the system (b) shows a zoomed-in view of (a) for only the six central grains and over a smaller time interval. Note that grains 18 and 23 act as effective walls for the four central soft grains, and the motion of these ``walls'' is evident in the kinetic energy density plot.}
\label{fig:fig1}
\end{figure}

Our goal is to engineer a simple system that can trap a significant fraction of the system energy. We propose that a small core of central grains, much softer than the main lattice, can do this, such as illustrated in Fig.~\ref{fig:fig1}a. In such a system, the soft central grains remain compressed for long times because it is energetically more favorable for the system to keep the soft grains compressed than to have a nonzero overlap between the harder grains. This introduces a restoring term to the force among the soft grains, and we can well-model the embedded system with a linearized force, possibly allowing us to tune the device's frequency response~\cite{Mohan2011}.

To excite this setup, we perturb the system as follows. We apply a single, discrete, symmetric perturbation from both ends, and this causes the two SWs to meet at the center of the chain, where they stay trapped for extended periods of time. Standing waves are then excited in the soft regions. 

Consider a one-dimensional chain of $N$ spherical grains between two fixed walls, under zero pre-compression. Initially the grains are barely touching, but any overlap between grains results in a Hertz~\cite{Hertz1882, Goldsmith1960, Sun2011} potential, $V(\delta_{i,i+1}) = a_{i,i+1} \delta_{i,i+1}^{5/2}$, where $\delta_{i,i+1} = R_i + R_{i+1} - (x_{i+1} - x_i)$ is the grain overlap with $x_i$ the absolute position, and 
\begin{equation}
a_{i,i+1} = \frac{2}{5 D_{i,i+1}}\sqrt{\frac{R_i R_{i+1}}{R_i + R_{i+1}}}, 
\label{eq:Fa}
\end{equation}
with $R_i$ the radius. The softness~\cite{Przedborski2015} is characterized by pre-factor $D_{i,i+1}$, which is related to the Young's modulus $Y$ and Poisson's ratio $\sigma$ of the grains as
\begin{equation} 
 D_{i,i+1} = \frac{3}{4}\left[ \frac{1-\sigma_i^2}{Y_i} + \frac{1-\sigma_{i+1}^2}{Y_{i+1}} \right ].
\label{eq:D}
\end{equation}
Fixed walls use $R_j \to \infty$, relaxing the condition that the wall must be flat. It is then possible to express the equations of motion for any grain in the chain as described previously~\cite{Przedborski2015}. We model our chains with soft grains centrally embedded in a longer chain of stainless steel grains of twice the radius (6~mm).
Choosing the central soft grains to be smaller than the hard grains creates an effective potential well at the center of the chain, since a sharp decrease in radius ensures complete transmission of momentum and energy, and partial reflection in the reverse case~\cite{Nesterenko1994,Nesterenko1995,Sen2001pro,Job2007,Sokolow2007}. 

To avoid SW formation in the center of the chain, the number of soft central grains should remain small (typically $3 \leq N_0 \leq 6$). For brevity, we report on the results only for $N_0$=3,4. The total number of grains is chosen to be $N$=39 or 40, to have an even number $(N-N_0)$ of stainless steel grains and preserve the symmetry. We then perturb the endmost grains at $t=0$ with a small initial velocity directed toward the center. 

We find that standing waves with distinct frequencies first form for systems whose central region corresponds to a $D$-value, see Eq.~\ref{eq:D}, which is at least two orders of magnitude larger than the host chain, i.e $D_s/D_h\sim10^2$, where $D_{s(h)}$ corresponds to the soft-soft(host-host) grain interaction. Further softening of the center slows energy propagation and leads to larger fractions of the system energy residing in central grains. This leads to beautiful harmonic behavior with clean, fast oscillations in the central region. For this reason, we present results for systems with central region $D_s\simeq 38$, consistent with material properties of rubber (stainless steel host chain corresponds to $D_h\simeq 0.007$). While these systems may be inevitably lossy, we focus on the short term dynamics and thus neglect dissipation in our studies.

% =============================================================================
% =============================================================================

Fig.~\ref{fig:fig1} shows the evolution of the $N=40$ system with $N_0=4$ soft grains by mapping the kinetic energy at each grain. Principal SWs approach the center region from the edges and are stopped at the edge of the soft grains for several ms before re-emerging.  Energy transmission at hard/soft grain boundaries, as well as reflections at the hard system boundaries, create many secondary solitary waves (SSWs), however the transition to the well known quasi-equilibrium phase is greatly delayed~\cite{Przedborski2015}. 
In Fig.~\ref{fig:fig1}a, we see that the emission of SWs from the soft region is slow compared to reflections at the walls, as SWs pick up speed until entirely free of the center grains. SSWs do not have constant speeds, traveling with erratic paths through the steel grains, indicating significant gaps have opened.

As the principal SW comes in the vicinity of the soft region, it slows down and remains nearly localized for a period of time before entering the soft region. This shows up as the dark horizontal lines seen bordering the soft region for long time spans in Fig.~\ref{fig:fig1}a. Once the SW enters the soft region, one can see several back and forth reflections before any energy is transmitted back into the hard grains. A closer look at the central region in Fig.~\ref{fig:fig1}b show these patterns to be similar to standing waves beating with quite distinct frequencies. Each grain oscillates with a similar high frequency while a lower frequency wave unites the movement of the central and edge soft grains.

% =============================================================================
\begin{figure}[htp]
\centering
%\framebox(150,125){TEXT}
\includegraphics[width=0.45\textwidth]{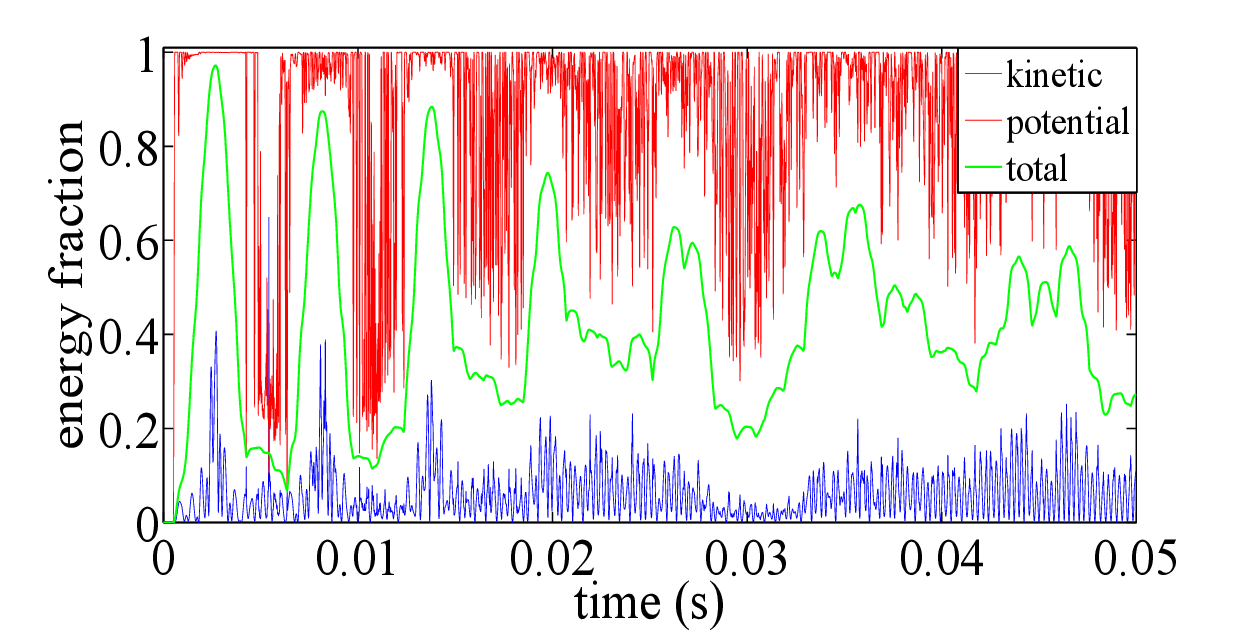}
\caption{Fraction of system energy (kinetic, potential and total energy) contained in the central soft grains in the $N=40$ system with four soft central grains for the first $0.05$s. From the virial theorem, the total system energy is split into 5/9 kinetic energy and 4/9 potential energy.}
\label{fig:fig2}
\end{figure}
% =============================================================================

The density of energy in the soft region remains high from the moment of initial compression by the first SW.   Fig.~\ref{fig:fig2} shows the fraction of the total system energy found in just the central region for the same configuration as Fig.~\ref{fig:fig1}, over time. The amount of stored compression (potential) energy in the soft grains fluctuates with the absorption and emission of SWs from the hard grains, however, on average more than $\sim80\%$ of the system's entire potential energy is found here. The soft grains remain compressed for $\gtrsim 1$~sec, the duration of the simulation, explaining the significant gaps among the hard grains that lead to the erratic paths of the SWs. 

\begin{figure}[htp]
\centering
%\framebox(150,125){TEXT}
\includegraphics[width=0.45\textwidth]{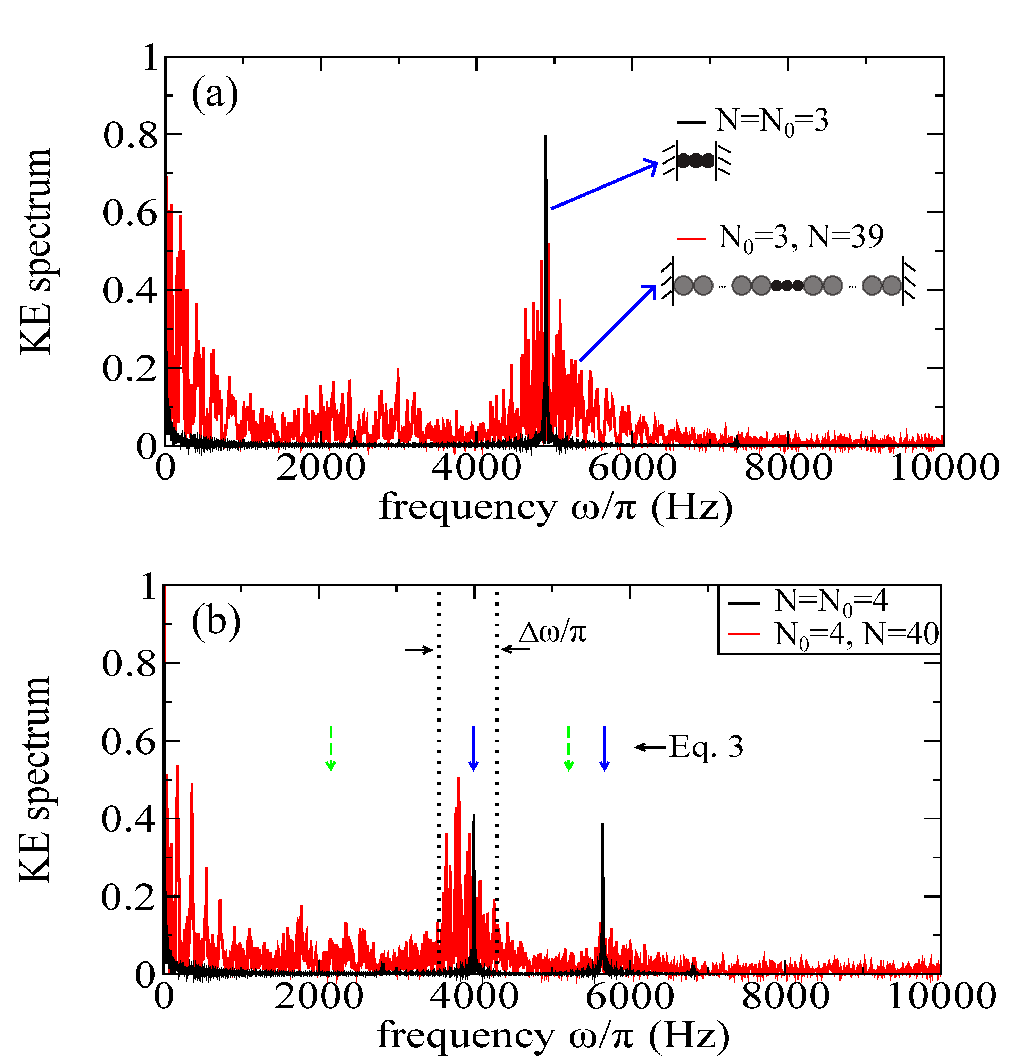}
\caption{Discrete cosine transforms of soft grain kinetic energy. In each DCT, we present both results of simulations between fixed walls and embedded in a steel grain matrix. (a) Three soft grains in an $N=39$ system and corresponding $N=3$ pre-compressed system for comparison. (b) Four central soft grains in an $N=40$ chain, and corresponding $N=4$ pre-compressed system. Arrows indicate the frequencies predicted by the harmonic model (dashed arrows indicate asymmetric modes which are not accessible when the system is perturbed symmetrically). The full width at half maximum (given by the dotted line) is predicted from the width of the distribution in spring constants.}
\label{fig:fig3}
\end{figure}

The discrete frequencies of the soft grain motion are clear in the oscillations of the kinetic energy fraction in Fig.~\ref{fig:fig2}. In Fig.~\ref{fig:fig3} we show the cosine spectrum of the total kinetic energy of soft grains. The spectra exhibit several low ($<700$~Hz) characteristic frequencies that correspond  to the periodic arrivals and departures of the SWs from the hard grains. From the kinetic energy density plots, the frequency of the initial SWs approaching the wall is $\sim 170$~Hz  for the four soft grain system. Note that there are also SSWs that frequently approach the soft grains, and these will contribute additional peaks to the DCT as well. 

More interestingly, the spectra also show a broad band at higher frequencies ($>3000$~Hz) that corresponds to the standing wave pattern of Fig.~\ref{fig:fig1}b. This band decreases in frequency with increasing number of soft grains, and a second band with frequencies $>5000$~Hz appear with four or more soft grains. We show that these frequencies are harmonic modes induced in the soft grains as a result of their large degree of compression, and that they contribute to the stored energy as well. 

% =============================================================================
% =============================================================================

To understand the origin of these frequencies, 
we also simulated a subset system comprised only of soft grains between fixed walls. We pre-compress the subset systems with an initial loading approximately equal to the average force experienced by soft grains in the embedded systems. In particular, we simulated $N=N_0=3,4$ grain systems with $F_0 = 3.5\times10^{-4}$~kN. As reference, the average forces at soft-grain interfaces in embedded systems were calculated to be $\langle F_{\mathrm{3}}\rangle = 3.7895\times10^{-4}$~kN, and $\langle F_{\mathrm{4}}\rangle = 3.1773\times10^{-4}$~kN.

The spectra of the small system kinetic energy is also shown in Fig.~\ref{fig:fig3} atop the spectra of the embedded systems. These show one or two very narrow bands whose frequencies correspond very well to the bands of the larger system. Pre-compression, and the resulting restoring force, is required to obtain these spectra; without pre-compression, frequency bands may not appear~\cite{Mohan2011}.

We can predict the frequency response of the embedded soft grains with a coupled harmonic oscillator model. For a four coupled mass-spring system between fixed walls, we obtain the following eigenfrequencies~\cite{Marion1988}:
\begin{eqnarray}
\label{eqn:freq}
\omega_{S\pm} &=& \left [\frac{1}{2m} \left ( k_w + 4k_g \pm \sqrt{k_w^2 - 4 k_w k_g + 8k_g^2}\right ) \right ]^{1/2}, \nonumber \\
\omega_{A\pm} &=& \left [ \frac{1}{2m} \left ( k_w + 2k_g \pm \sqrt{k_w^2 + 4 k_g^2} \right ) \right ]^{1/2}, \end{eqnarray}
where subscript $S$ corresponds to symmetric modes in which motion of the central grains is mirror reflected about the center of the chain, and subscript $A$ to anti-symmetric modes in which there is no mirror reflection symmetry. In this last expression,
$k_w$ is the grain-wall spring constant, and $k_g$ is the grain-grain spring constant, noting that the grain-wall connection corresponds to the soft-hard grain interface for the embedded system. The angular frequencies, $\omega$, are related to the frequencies measured in kinetic energy spectra via $\omega / \pi = f$, since the positions and velocities of the grains oscillate as $\cos(\pi f t)$ while kinetic energy oscillates as the velocity squared, or $\cos(2 \pi f t)$. 

The average force between the soft grains and the soft grains and walls is the same. We find the effective spring constants for the \emph{grain-grain} and \emph{grain-wall} forces at the pre-determined average force magnitude using the force vs. $\delta$ plots (not shown). In particular, the slopes of these plots at the average force magnitude give $k_g$ and $k_w$, respectively. The fluctuations about the average force are wide. If we take the value of the slope of the grain-wall force at these limits of one standard deviation from the average, we obtain extrema of $k_w$. Likewise we can find extrema of $k_g$.

Using Eq.~\ref{eqn:freq}, the average values of $k_w$ and $k_g$ predict the center frequency and the two extrema predict the bandwidth of the kinetic energy spectrum. These results are indicated by the arrows in Fig.~\ref{fig:fig3}b, where the solid arrows indicate the predicted frequencies of the symmetric modes $\omega_{S\pm}$, and the dashed arrows indicate the asymmetric modes $\omega_{A\pm}$. The frequencies of both the embedded and small test systems are precisely predicted by the harmonic model. Furthermore, the bandwidth of $\omega_{S-}$ is indicated by the vertical dashed lines, and does encompass a majority of the principal frequency band. Thus the broad bandwidth of the harmonic motion of the soft central region is due to changing compression forces caused by its variable boundaries. 

These values of $k_w$ and $k_g$ also agree with the $N=N_0=4$ soft system. The inter-grain force is $F_{ij} = \frac{5}{2} a \delta_{ij}^{3/2}$, which in the limit that the initial pre-compression in the model Hertzian system is large, i.e. $\delta_0 \gg u_{i-1} - u_i$, can be Taylor expanded to be $F_{ij} =\frac{15}{4} a \delta_0^{1/2} \left ( u_{i-1} - u_i \right )$.  The effective spring constant is then given by $k_{\mathrm{eff}} = \frac{15}{4} a \delta_0^{1/2}$. Using the values for $a$ and $\delta_0$ for rubber with steel walls, we find $k_w = 0.02141$~kN/mm and $k_g = 0.010707$~kN/mm for the $N=N_0=4$ soft system, which is in agreement with linear fits to the force vs. displacement curve described above for the $N_0=4,N=40$ system.

Due to how the system is perturbed, the resulting motion of the soft grains is a combination of only the symmetric modes with corresponding eigenvectors
\begin{equation}
\vec{\omega}_{S\pm} = \frac{1}{2\sqrt{m}} \left ( 1, \mp 1, \pm 1, -1 \right )^{T}, 
\end{equation}
where we have used the fact that $k_w \approx 2k_g$ in the last expressions. The asymmetric modes can be populated only with asymmetric perturbations, e.g. single edge perturbations or unequal perturbations from both ends (data not shown). It can be shown that the total (kinetic plus potential) energy of each \emph{harmonic} mode is exactly the same and equal to $\frac{1}{2}mv_0^2$, where $v_0$ is the initial velocity of the first mass. 

For the embedded system, $v_0$ is ill-defined but is dependent on the instantaneous boundary conditions of the steel grains containing the soft grains. The arrival or departure of every SW transmits an additional acceleration perturbation to the soft grains, which may be positive or negative depending on whether the boundaries of the soft grains are  contracting or expanding at that moment. Thus the amount of energy in the harmonic modes can vary, however, this is just a small fraction of the total system energy. Since the soft grains are much softer and have less mass than the steel grains, most of the system's energy is contained in the form of stored potential energy due to the initial compression of the soft grains.

% =============================================================================

\begin{figure}[ht]
\centering
%\framebox(150,125){TEXT}
\includegraphics[width=0.48\textwidth]{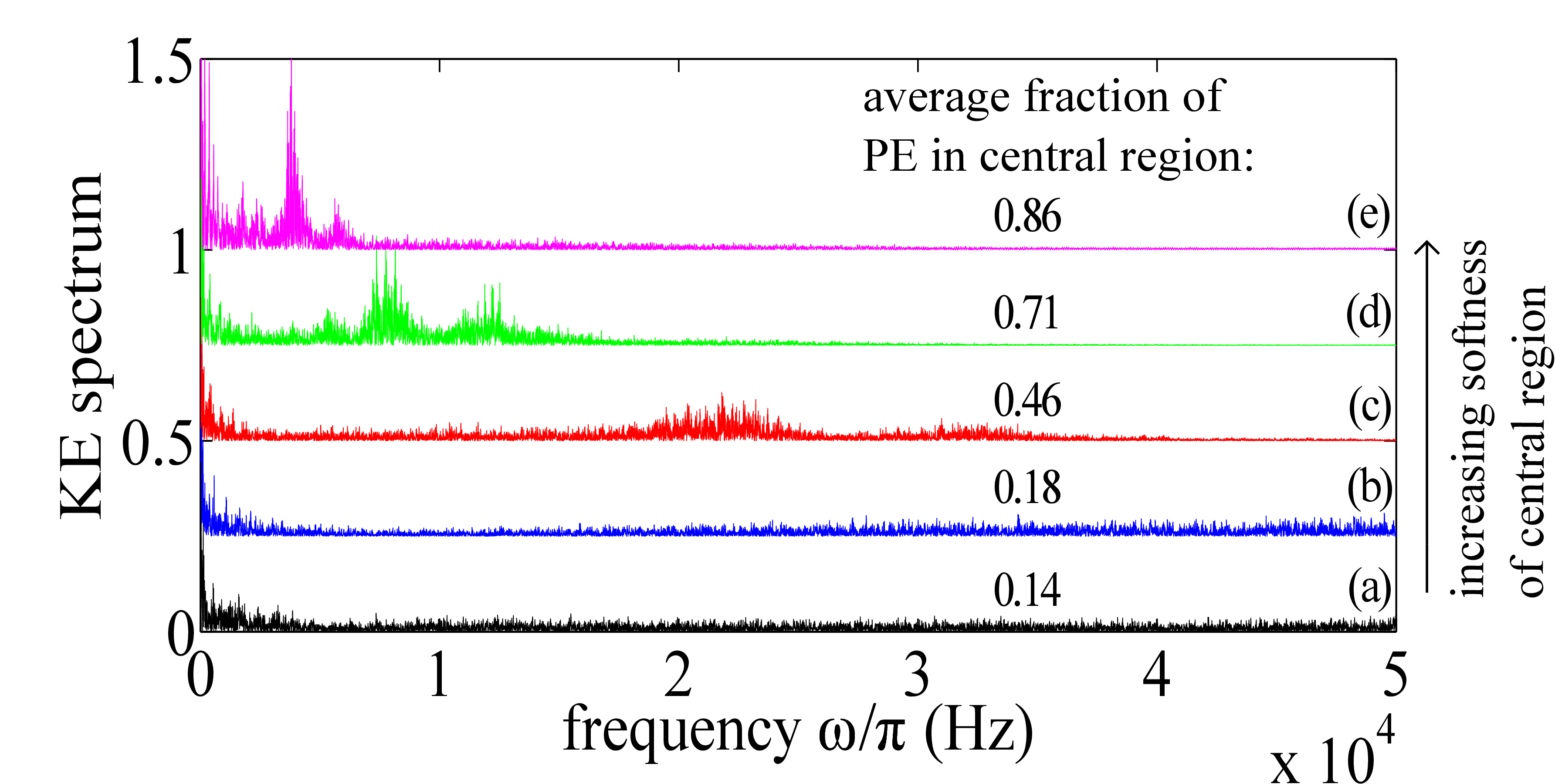}
\caption{Comparison of kinetic energy spectra for $N_0=4, N=40$ chains with varying central grain softness; (a) stainless steel control experiment, (b) Pyrex, (c) PVC, (d) PTFE, (e) rubber. The average fraction of the total system potential energy (PE) that is contained in the central region is also shown. See Ref. \onlinecite{Przedborski2015} for a table of material values.}
\label{fig:fig4}
\end{figure}

For the system potential energy to be located almost exclusively in the central region, the central grains must be sufficiently softer than the matrix. We tested a range of materials with $D_{s}\approx0.007\rightarrow38.0$ (larger is softer). 
Fig.~\ref{fig:fig4} illustrates that pseudo-harmonic frequencies appear only when the fraction of system potential energy in the central region exceeds $\sim40$\%, which happens when $D_s/D_h\sim10^2$.
These frequencies shift lower with increasing softness, reflecting a softening harmonic spring constant $k_{\mathrm{eff}}$ with increasing $D$. Observe that $k_{\mathrm{eff}}\propto a\delta_0^{1/2}$, where $a\propto D^{-1}$ faster than $\delta_0$ increases with $D$.

We have demonstrated a simple way of localizing energy to a small region in an initially uncompressed Hertz chain. What is unique about this system is that one may engineer a means of tapping into the predictable frequencies and energy of the long-lived harmonic modes that develop in the central grains, even if the perturbations of the surrounding harder matrix are unpredictable in time or amplitude. 
Our system converts symmetric SWs to oscillations. We also tested asymmetric perturbations (not shown), and the central results remain unchanged. 

Since we are considering relatively short times, dissipation is not important in our study.
Furthermore, a dissipation-free version of this setup may be of interest in the up-conversion of low to high frequencies. 
Since our system is capable of converting delta perturbations to oscillations with frequencies in the kHz regime, it is reasonable to think that one may achieve up to a three decade increase in frequency in systems with no dissipation. In addition, this device can be generalized into three dimensions by the assembly of one dimensional columns into a lattice. Such a system would exceed the energy absorption of the device recently described by Breindel et al.~\cite{Breindel2011} 

% =============================================================================
% =============================================================================

% ================================================================================
\begin{acknowledgments}
This work was supported by a Vanier Canada Graduate Scholarship from the Natural Sciences and Engineering Research Council. S.S. thanks US Army Research Office for partial support of this research through a Short Term Innovative Research Grant.
\end{acknowledgments}
% ================================================================================

% ================================================================================
\bibliography{Localizing_energy}

\end{document}